# Compound Effect of Alfvén Waves and Ion-cyclotron Waves on Heating/Acceleration of Minor Ions via the Pickup Process

C. B. Wang[1*], Bin Wang[2], L. C. Lee[3]

[1]CAS Key Laboratory of Geospace Environment, School of Earth and Space Sciences, University of Science and Technology of China, Hefei, 230026, China.

[2] Beijing Institute of Tracing and Telecommunications Technology of China, Beijing, China

[3]Institute of Earth Sciences, Academia Sinica, Taipei, 11529, Taiwan

*Corresponding author, email address: cbwang@ustc.edu.cn

**Abstract**

A scenario is proposed to explain the preferential heating of minor ions and differential streaming velocity between minor ions and protons observed in the solar corona and in the solar wind. It is demonstrated by test particle simulations that minor ions can be nearly fully picked up by intrinsic Alfvén-cyclotron waves observed in the solar wind based on the observed wave energy density. Both high frequency ion-cyclotron waves and low frequency Alfvén waves play crucial roles in the pickup process. A minor ion can first gain a high magnetic moment through the resonant wave-particle interaction with ion-cyclotron waves, and then this ion with a large magnetic moment can be trapped by magnetic mirror-like field structures in the presence of the lower-frequency Alfvén waves. As a result, the ion is picked up by these Alfvén-cyclotron waves. However, minor ions can only be partially picked up in the corona due to low wave energy density and low plasma beta. During the pickup process, minor ions are stochastically heated and accelerated by Alfvén-cyclotron waves so that they are hotter and flow faster than protons. The compound effect of Alfvén waves and ion-cyclotron waves is important on the heating and acceleration of minor ions. The kinetic properties of minor ions from simulation results are generally consistent with *in situ* and remote features observed in the solar wind and solar corona.

# 1. Introduction

The heating and acceleration of ions in the solar corona and solar wind has been the subject of extensive scientific research for several decades. Minor ions, namely, alpha particles and other heavy ions which are heavier than helium, carry much information on the collisionless solar plasma. The minor ions can be used as diagnostic tracers to probe the solar atmosphere. Many remote and *in situ* observations have paid attentions to the kinetic properties of these minor ions. These observational results, summarized in the following, provide very important information and stringent constraints on the heating and accelerating processes of solar corona and solar wind.

1. In the corona hole at a few solar radius, spectroscopic remote sensing observations of minor ion emission lines by SOHO Ultraviolet Coronagraph Spectrometer (UVCS) show that minor ions such as $O^{5+}$ have a strong temperature anisotropy ($T_\perp / T_\parallel > 10$). Its perpendicular temperature are greater than $10^8$ K, which is much higher than the proton temperature $\approx 3 \times 10^6$ K. The flow speed of $O^{5+}$ ions can exceed that of protons as much as 200-300 km/s [Kohl et al. 1997, 1998; Li et al., 1998; Cranmer et al., 1999a, 1999b; Kohl et al., 2006], which is about 0.1-0.3 times of the local Alfvén speed that is about 1000-2000 km/s in the upper corona.

2. In the solar wind between 0.3 AU to 1 AU, satellite *in situ* observations of the solar wind indicate minor ions also flow faster than protons with a relative speed equal to or less than the local Alfvén speed. The typical condition for the total temperature anisotropy of alpha particles is $T_{\perp\alpha} < T_{\parallel\alpha}$ [Marsch et al., 1982b; Reisenfeld et al., 2001; Gary et al., 2005]. Nevertheless, an opposite anisotropy with $T_{\perp\alpha} / T_{\parallel\alpha} > 1$ is observed at times [Neugebauer et al., 2001; Reisenfeld et al., 2001; Gary et al., 2002; Bourouaine et al., 2011a]. The helium temperature anisotropy ($T_\perp / T_\parallel$) rises with decreasing of the flow speed difference between helium ions and protons [Gary et al. 2005; Kasper et al. 2008]. The velocity distribution of helium ions occasionally shows a double-peak structure in the solar wind [Marsch et al., 1982b].

3. In the interplanetary space beyond 1 AU, the Ulysses-SWICS data show that different minor species can have approximately equal thermal speeds in the fast solar wind. Their bulk speeds are also with similar average values and faster than that of protons by about the local Alfvén speed [von Steiger et al., 1995; von Steiger & Zurbuchen, 2006].

A number of theories have been proposed on the preferential heating and acceleration of minor ions in the solar corona and solar wind. According to the classification of Cranmer [2009], there seems to be two broad classes of physics-based models. One is the reconnection/loop-opening (RLO) models. In these models, the solar corona and wind are assumed to be influenced by impulsive bursts in the lower atmosphere, such as the magnetic reconnection between closed, loop-like magnetic flux systems. For example, Lee & Wu (2000) and Lee (2001) suggested that ions can be heated and accelerated by subcritical fast shocks. These shocks are generated by small-scale reconnection events at the solar surface to propagate outward into the extended corona. Another is the wave/turbulence-driven (WTD) models. This is mainly based on observations that intrinsic large-amplitude Alfvén-like waves are present in the interplanetary medium [Belcher & Davis 1971; Smith et al., 2006; Wang et al., 2012]. Alfvén wave-like fluctuations have also been measured remotely in the chromosphere and corona [De Pontieu et al., 2007;Tomczyk et al., 2007; McIntosh et al., 2011].It is generally expected that through the resonant wave-particle interactions, Alfvén-cyclotron waves play an important role for the heating and acceleration of solar corona and solar wind [Hollweg & Turner, 1978; Marsch et al., 1982a; Cranmer, 2001; Tu and Marsch, 2001; Gary et al., 2001; Vocks and Marsch, 2002; Hollweg and Isenberg, 2002; Marsch et al., 2006;

Araneda et al 2008; Isenberg and Vasquez, 2011].

However, there are theoretical difficulties with the application of the ion cyclotron mechanism, and its role is not yet fully understood [Isenberg and Vasquez, 2007; Ofman, 2010].The frequency of waves generated by the convection-driven jostling of magnetic flux tubes in the photosphere is very low, which cannot become resonant with ions and heat them. Moreover, the fluctuating power of high-frequency Alfvén-cyclotron waves is also much lower than that of the low-frequency Alfvén waves in the observed power spectrum of the magnetic fluctuations in the solar wind beyond 0.3 AU [Tu & Marsch, 1995; Smith et al., 2006; Bourouaine et al., 2012]. To overcome this obstacle, it has been suggested that high frequency ion-cyclotron waves can be generated from low frequency Alfvén waves through MHD turbulent cascade [e. g., Tu et al., 1984]. On the other hand, Axford & McKenzie (1992) suggested that high frequency waves can be excited directly by microflares in the choromosphere.

The low-frequency Alfvén waves and kinetic Alfvén waves have attracted extensive attentions recently. Several different theoretical approaches, which are not based on the cyclotron resonant interaction, have been suggested. Firstly, it has been argued that in low-beta plasma condition, due to the pitch-angle scattering of ions, the ion temperature anisotropy could be caused by Alfvénic fluctuations with frequencies well below the local ion-cyclotron frequency [Wang et al, 2006; Wu & Yoon 2007; Li et al., 2007; Bourouaine et al., 2008; Wang & Wu, 2009; Wang & Wang, 2009; Nariyuki et al., 2010; Verscharen & Marsch, 2011; Liu et al., 2013; Dong, 2014]. The energy gain for the ions is proportional to $\delta B_w^2$, where $\delta B_w$ is the average magnetic field wave amplitude. The second approach imposes a somewhat different concept of heating by emphasizing stochasticity of particle motion. When the wave amplitude exceeds certain threshold value, the motion of particle may change from regular to stochastic due to high order nonlinear resonance or the kinetic effect of finite ion lamor-radius [Lin & Lee, 1991; Chen et al., 2001; White et al., 2002; Voitenko and Goossens, 2004; Wu and Yang, 2007; Lv et al., 2007; Guo et al., 2008; Chandran et al., 2010; Wang et al. 2011].

The heating rate and acceleration of waves on different species of ions have been analyzed from models based on the cyclotron resonant interactions [Dusenbery and Hollweg, 1981; Marsch et al., 1982a; Isenberg and Hollweg, 1983; Hu and Habbal, 1999; Cranmer, 2001; Marsch and Tu, 2001; Vocks and Marsch, 2002; Hollweg and Isenberg, 2002; Marsch et al., 2006; Isenberg and Vasquez, 2011] and those not based on the cyclotron resonant interactions [e.g. Voitenko and Goossens, 2004; Wu and Yang, 2007; Chandran et al., 2010, 2013]. All models indicate that heavy ions can be preferentially heated and accelerated with respect to protons. The results appear to be consistent with the recently observed kinetic properties of ions in the solar wind [for example, Kasper et al., 2013; Chandran et al., 2013]. However, there is a broad spectrum of Alfvén-cyclotron waves in the solar wind from observations, which include both low frequency waves (LFWs) (far below the ion gyro-frequency, Alfvén waves) and high frequency waves (HFWs) (near the ion gyro-frequency, ion-cyclotron waves). In this paper, we examine the different roles of the LFWs and HFWs in the heating and acceleration of ions in the solar corona and solar wind.

In a previous paper [Wang et al. 2011], we discussed the stochastic heating and acceleration of minor ions by low-frequency obliquely propagating Alfvén waves in the solar wind. The asymptotic status of the stochastic heating and acceleration was emphasized in the paper. It was found that when the wave amplitude exceeds some threshold condition for stochasticity, the time-asymptotic kinetic temperature associated with the minor ions becomes independent of the wave amplitude, and the temperature always approaches a value dictated by the Alfvén speed. During the course of the heating process the minor ions gain a net average parallel speed approximately equal to the Alfvén speed in the plasma frame. The physical mechanism for the asymptotically heating is the pickup process that involves the formation of spherical shell velocity distribution function due to the pitch-angle scattering. This stochastic heating process has a potential application to explain the observational properties of minor ions in the solar wind and solar corona. However, the time evolution of the ion velocity distribution was not discussed in that paper. Whether the threshold condition for stochasticity can be satisfied in solar corona and solar wind is still an open question. Recently, Bourouaine et al. (2011a, 2011b) showed observational results on both the kinetic characters of alpha particles and the normalized power of magnetic transverse fluctuations in solar wind at a heliocentric distance of about 0.7 AU from the past Helios mission in 1976. These observations give us an opportunity to examine whether the stochastic mechanism is operating in the solar wind.

In this paper, we will study the compound effects of Alfvén waves and ion-cyclotron waves on the pickup of minor ions by a spectrum of intrinsic Alfvén-cyclotron waves, and a simple scenario is proposed for the preferential heating and acceleration of minor ions in the solar corona and solar wind. The physical processes and the different roles of LFWs and HFWs for the ion pickup are investigated in detail. One of our most important finds is that both LFWs and HFWs play key roles for ion pickup by turbulent Alfvén-cyclotron waves in the solar wind. HFWs are important because they can randomize the ion orbit and destroy the conservation of ion magnetic moment through resonant wave-particle interactions. LFWs are important because they can reflect and trap ions in the Alfvén frame due to magnetic mirror force since their amplitudes are much larger. The compound effect of Alfvén waves and ion-cyclotron waves leads to the stochastic heating and acceleration of minor ions so that they are hotter and flow faster than protons. However, minor ions can only be partially picked up in the corona due to the low wave energy density and plasma beta value. The temperature anisotropy and velocity distribution of minor ions at different stages of pickup is also discussed and compared with observations in solar wind and solar corona.

## 2. Basic Assumption and Physical Model

We assume that intrinsic Alfvén-cyclotron waves pervade in the solar corona and solar wind, which propagate outward from the Sun. Because the abundance of minor ions is low, we also assume that even if there may be waves excited locally by minor ions, their energy density is low compared with that of intrinsic waves. In other words, we assume that the presence of minor ions do not have any significant impact on the intrinsic turbulence. Consequently, we adopt the test particle simulations for simplicity and clarify. Let the ambient magnetic field $\mathbf{B}_0$ be in the $z$ direction. Without loss of generality, a spectrum of linearly polarized incoherent Alfvén-cyclotron waves with parallel propagation along the ambient magnetic field is considered. The magnetic and electric wave fields in plasma frame can be expressed as

$$\delta \mathbf{B}_\mathbf{w} = \sum_{j=1}^{N} B_j(t) \sin \psi_j \hat{\mathbf{y}}, \tag{1a}$$

$$\delta \mathbf{E}_w = \sum_{j=1}^{N} \frac{\omega_j}{k_z} B_j(t) \sin \psi_j \hat{\mathbf{x}} \tag{1b}$$

Where $\psi_j = \omega_j t - k_z z + \varphi_j$; $k_z$ is the wave number parallel to the ambient magnetic field; $\varphi_j$ and $B_j$ are the random phase constant and amplitude for wave mode $j$; $\omega_j$ is the wave frequency; $\hat{\mathbf{x}}$ and $\hat{\mathbf{y}}$ are unit vectors along the $x$ and $y$ direction. The wave dispersion relation is

$$\frac{c^2 k_z^2}{\omega_j^2} = 1 + \frac{c^2}{v_A^2} \frac{1}{1 - \omega_j^2 / \Omega_p^2},$$

where $V_A = B_0 / \sqrt{\mu_0 \rho_0}$ is the Alfvén speed, and the dispersion for different wave mode is included. The amplitude of each wave mode satisfies a power-law relation with a spectral index $-\alpha$, $B_j^2 \propto \omega_j^{-\alpha}$, and $\alpha$ is chosen as 5/3 that is a generally accepted value for the power spectrum of magnetic fluctuations found in the solar wind [e.g., Tu & Marsch 1995; Bourouaine et al., 2012]. The minor ions are adopted as oxygen $O^{5+}$ whose dynamics is governed by

$$m_{O^{5+}} \frac{d\mathbf{v}}{dt} = q_{O^{5+}} \left[ \mathbf{E}_w + \mathbf{v} \times (\mathbf{B}_0 + \delta \mathbf{B}_w) \right], \quad \frac{d\mathbf{r}}{dt} = \mathbf{v} \quad (2)$$

Where $m_{O^{5+}}$ and $q_{O^{5+}}$ are the mass and electric charge of $O^{5+}$. The equations of motion are solved with Bulirsch-Stoer algorithm [Stoer & Bulirsch, 1980] in a frame moving with the Alfvén speed (Alfvén frame). In the Alfvén frame, the wave electric field is very weak due to weak dispersion, and hence the ion kinetic energy is nearly conserved in the simulation. We discretize the Alfvén-cyclotron wave spectrum as follows: $\log \omega_j = \log \omega_1 + (j-1)\Delta\omega$, $(j = 1, 2, \cdots, N)$, where $\Delta\omega = (\log \omega_N - \log \omega_1)/(N-1)$, $N$ is the total number of wave modes. To avoid the strong initial-pitch-angle scattering effect [Wang & Wang, 2009], we let that the amplitude of each wave mode changes gradually with time from an initially small value to a finite amplitude such that the wave energy density $\delta B^2(t)/B_0^2 = \sum_{j=1}^{N} B_j^2(t)/2 = \varepsilon(t)$, and

$$\varepsilon(t) = \begin{cases} \varepsilon_0 e^{-(t-\tau)^2/(\Delta\tau)^2}, & t < \tau \\ \varepsilon_0, & t \geq \tau \end{cases}, \quad (3)$$

where $\varepsilon_0$ is the final wave energy density normalized with the ambient magnetic field energy. We choose $\tau = 200\Omega_p^{-1}$ and $\Delta\tau = 50\Omega_p^{-1}$, where $\Omega_p = eB_0/m_p c$ is the proton cyclotron frequency. The physical consideration is that minor ions as well as waves are co-moving with the solar wind. When Alfvén waves propagate outward from the lower corona (where the plasma beta value $\beta < 0.01$) to the interplanetary region (where $\beta \geq 0.1$), the wave energy density normalized with the local ambient magnetic field would increase.

The velocity and time are normalized with respect to the Alfvén speed $v_A$ and $\Omega_p^{-1}$, respectively. The time step is $\Omega_p \Delta t = 0.001$. There are $10^4$ test particles, which are initially distributed at random during the time interval $\Omega_p t = [0, 2\pi]$ and along the spatial range $z\Omega_p / V_A = [0, 3000]$. Their initial velocities are assumed to have a Maxwellian distribution with thermal speed $v_{TO^{5+}}$.

## 3. Mechanism for the Pickup of Minor Ions by Alfvén-Cyclotron Waves in Solar Wind

As mentioned in the introduction, there are mainly two scenarios on studying Alfvén-wave-particle interaction in solar corona and solar wind in the literatures. One emphasizes the resonant wave-particle interactions with high frequency ion-cyclotron waves. The other emphasizes the nonlinear stochastic heating and acceleration by large amplitude low frequency Alfvén waves which cannot resonate with ions based on the linear theory. A concern for the former is that the energy density of HFWs, which can resonate with ions, are generally much low. A question for the later is whether the amplitude of LFWs is really large enough to exceed the threshold for stochasticity in solar corona and wind. We would like to demonstrate in this section that minor ions can be nearly fully picked up by the intrinsic Alfvén-cyclotron waves in solar wind, in which both HFWs and LFWs play key roles.

To clarifying the role of HFWs and LFWs, three simulation cases are studied in this section. In Case 1, including both LFWs and HFWs or the whole wave spectrum, the simulation parameters are $\omega_1 = 0.01\Omega_p$, $\omega_N = 0.4\Omega_p$, $N = 51$, and $\varepsilon_0 = 0.1$. The wave energy ratio $\varepsilon_0 = \delta B^2 / B_0^2 \approx 0.1$ is based on observations in the solar wind. Bourouaine et al. [2011b] investigated the integrated wave power observed by Helios at 0.7 AU in 1976 over the wave number range, $(0.01-1)k_p$ in the plasma frame, where $k_p = \Omega_p / V_A$ is the proton inertial length. It is found that these waves are mainly transverse and essentially incompressible. The averaged value of the normalized wave power $\delta B^2 / B_0^2$ is about 0.1-0.2 in the fast solar wind where the collisional age is less than 0.1. The observed parallel plasma beta is comparatively small and mainly varies between 0.1 and 0.7. In Case 2, including only HFWs that can resonant with $O^{5+}$ ions, the wave modes of number $j = 42, 43, ..., 51$ in Case 1 are used, which cover the frequency range $[0.2, 0.4]\Omega_p$. In Case 3, including only LFWs that cannot resonate with $O^{5+}$ ions in general, the wave modes of number $j = 1, 2, ..., 35$ in Case 1 are employed, which cover the frequency range $[0.01, 0.12]\Omega_p$. Each wave mode j in Case 2 and Case 3 has the same amplitude as its corresponding value with same number j in Case 1. The HFWs and LFWs used in Case 2 and Case 3 occupy about 1% and 90% energy of the whole spectrum in Case 1, respectively. Protons and minor ions are assumed to have the same initial thermal speed, $v_T = v_{TO^{5+}}$. In Cases 1-3, we set $v_T = v_{TO^{5+}} = 0.3V_A$, corresponding to $\beta \approx 0.1$. Parameters for different simulation cases are summarized in Table 1.

Figure 1 is the velocity scatter-plot of $O^{5+}$ ion in the $v_z$ versus $v_\perp$ phase plane for Case 1 (top panel), Case 2 (medium panel) and Case 3 (bottom panel) in the plasma frame at different simulation times. For Case 1, there is strong stochastic heating and acceleration, $O^{5+}$ ions can be nearly fully picked up by the waves through forming a spherical shell-like distribution at the end of simulation. This spherical shell-like distribution is the result of ion energy nearly conserving in the Alfvén frame, which will be discussed in more detail in the next section. The center of the spherical shell is located at the Alfvén speed $v_z = V_A$. In Case 2, although $O^{5+}$ ions can be heated (mainly in the perpendicular direction) due to the resonant wave-particle interaction, but very few ions can be pitch-angle scattered to the right half-side of the spherical shell. The accelerated average parallel velocity can hardly greater than $0.4V_A$. In Case 3, there are even fewer ions scattered to the right half-side of the spherical shell at the end of simulation. Clearly, the wave amplitude does not exceed the threshold value of stochasticity for most ions in Case 3. The increase of the ion kinetic energy in Case 3 is mainly due to the non-resonant pseudoheating process discussed in Wu et al [2007] and Wang & Wu (2009).

These results imply that both HFWs and LFWs are important for the pickup of minor ions by Alfvén-cyclotron waves in solar wind. To understand the different roles that HFWs and LFWs play during the pickup process, we track the moving trajectories of individual particles at different times in the three simulation cases. Figure 2 shows variations of the magnetic moment and position $z$ of an ion with time in the Alfvén frame for the above three simulation cases. The results for Cases 1, 2 and 3 are represented by the red, green and blue lines, respectively. Each ion has the same initial velocity and position in all three simulation cases. In Figure 3, the magnetic moment, $\mu = m_\alpha v_\perp^2 / 2|\mathbf{B}_t|^2$, is defined with respect to the local total magnetic field $\mathbf{B}_t = \mathbf{B}_0 + \delta\mathbf{B}_w(z,t)$, where $\delta\mathbf{B}_w(z,t)$ is the wave magnetic field experienced by the particle at position $z$ and time $t$. It is found that the ion magnetic moment $\mu$ changes greatly with time in both Case 1 and Case 2, while its variation with time is relatively small in Case 3. The reason is that there are HFWs in both Case 1 and Case 2, the magnetic moment $\mu$ of an ion is not conserved due to the resonant wave-particle interaction between ion and HFWs. However, there are only LFWs in Case 3, whose amplitude is not large enough to have a strong nonlinear stochastic heating for this ion. The slow variation of the ion magnetic moment $\mu$ with time in Case 3 is mainly due to the small dispersion of different wave modes and weak stochasticity for this ion.

In Figure 2, the ion position z decreases monotonically with the increase of time in Cases 2 and 3. This means that the ion is always streaming away along the direction anti-parallel to ambient magnetic field in the Alfvén frame. One the other hand, the same ion change moving directions from anti-parallel direction to parallel direction, and vice versa, for several times in Case 1. This means that this ion is trapped in the wave field. In other word this ion is picked up by waves in Case 1. The physical reason is that the ion with a relatively large $\mu$ can be intermittently bounced backward and forward, or from left-side (right-side) to the right-side (left-side) of the spherical shell in Figure 1, by the large amplitude LFW field due to magnetic mirror force in Case 1 in the Alfvén frame.

This reflecting and trapping process can be seen more clearly from Figure 3, which illustrates the trajectory of this same ion along the *z* axis (dash line) and the wave field strength (solid line) at three time intervals in Case 1. In the Alfvén frame, the amplitude of turbulent LFWs changes slowly in space. The wave field strength is weak in some areas, while it can have relatively large values at some other positions. Thus, a number of magnetic mirror-like field structures can be formed in the Alfvén frame. In some conditions, the ion can be bounced back at the mirror point, or, even can be trapped by these mirror-like wave field structures as shown in Figure 3. It is found that this ion is reflected or bounced back 3 times in the region $z \approx [-1600, -1400] V_A / \Omega_p$ (bottom panel in Figure 3), once at the position $z \approx -50 V_A / \Omega_p$ (medium panel), and 8 times in the region $z \approx [-5400, -5250] V_A / \Omega_p$ (top panel). This ion experiences a large amplitude wave magnetic field at each time when the ion is bounced back. The times of ion reflection are also indicated by red short vertical bars in Figure 2 for Case 1. One can see from Figure 2 that the magnetic moment $\mu$ of the ion has a high value, generally greater than $1.3 m_\alpha V_A^2 / 2B_0$, at the time when the ion is bounced back or trapped by mirror-like field structures. This is not a surprising result. Note that the magnetic moment shown in figure 3 is defined by the local total magnetic field, and the ion energy is nearly conserved in the Alfvén frame. The higher the magnetic moment is, the larger the velocity component perpendicular to local magnetic field, and the smaller the parallel velocity component. Obviously, an ion with small parallel velocity is more easily to be bounced back when the ion moves in a region with gradually increasing strength of magnetic field. Of course, if the variation of magnetic field strength is too small, the ion cannot be bounced back even if its magnetic moment is high as in the Case 2 where no large amplitude LFWs exists.

In short, from the above results, we consider that both LFWs and HFWs play important roles for the pickup of minor ions in solar wind. The basic physical process can be described as following. A minor ion can obtain a high magnetic moment through the resonant wave-particle interaction with HFWs, because HFWs mainly heat minor ion in the direction perpendicular to the local magnetic field. Then, an ion with a large magnetic moment can be intermittently bounced backward and forward or trapped by magnetic mirror-like field structures formed by the large amplitude LFWs in the Alfvén frame. When an ion is trapped in the wave field, it will co-move with the waves. In other words, the ion is picked up by these waves.

Finally, we like to point out that if the amplitude of LFWs is large enough, the nonlinearly higher order resonant interactions between ion and LFWs would become significant. The magnetic moment of a minor ion can also increase to enough large value to let the ion being picked up even without HFWs (Wang et al., 2011). However, it seems that this is not easy to occur for the observed strength of LFWs in the solar corona and solar wind.

# 4. Kinetic Properties of Minor Ions During the Pickup Process in the Solar Wind

During the pickup process, minor ions will be stochastically heated and accelerated by the Alfvén-cyclotron waves. In this section, we will give some discussions on the kinetic properties of minor ions observed in the solar wind based on the pickup process described in the above section. As mentioned in the introduction, the kinetic properties of minor ions have been the subject of extensive studies from both theories and *in situ* observations. The properties of minor ions that distinguish them from protons can help us understand the mechanisms for the heating and acceleration of solar corona and solar wind.

Due to the low abundance of minor ions, the three dimensional velocity distribution of minor ions has not been measured in situ except helium. However, a common feature of observations for both helium and other minor ions is that they are hotter than protons, and flow faster than protons with a relative speed less or equal to the Alfvén speed. We may expect that other minor ions have the similar distribution of velocity as helium ions. The helium abundance may vary from 1 to 6 percent in usual solar wind conditions [Kasper et al., 2007]. Rigorously, we would like to note that the simulation results discussed in this paper need further studied for helium ion when its abundance is relatively large.

To study the effect of plasma beta, another simulation, Case 4 is simulated with initial thermal speed $v_T = v_{TO^{5+}} = 0.7 V_A$ (the corresponding plasma beta $\beta \approx 0.5$). The other simulation parameters in Case 4 are as same as that in Case 1 ($\beta \approx 0.1$).

## 4.1 Velocity distribution and formation of double streaming peaks.

The normalized distributions of ion velocities in plasma frame at different time are shown in

Figure 4. The top (bottom) panel is for Case 1 (Case4), and the left (right) column is for the $v_x$ ($v_z$) component velocity. The black solid line, red dot line, green dash line, and blue dot-dash line represent the result at time $\Omega_p t = 0, 2500, 10000, 20000$, respectively. The evolution of the $v_y$ component velocity distribution (not shown) is similar to the $v_x$ component. The broadening of velocity distribution with increasing time represents the continual stochastic heating of minor ions. The movement of the center of $v_z$ distribution from left to right indicates the acceleration of ions in the parallel direction, or the gradual ion pickup by Alfvén-cyclotron waves.

A very interesting character of the velocity distribution is the double streaming peaks and the asymmetry along the direction parallel to the ambient magnetic field. One can see from the right column of Figure 4 that as minor ions are gradually picked up by Alfvén waves, the peak of parallel velocity is moving in the wave propagating direction, and a second peak starts to form. The distance between the two peaks is approximately equal to the Alfvén speed. This double peak distribution of velocity has been observed for helium ions in the solar wind by Helios. In the classic paper by Marsch et al. [1982b], it is found that '*the helium ion double peak distributions occur in the solar wind and can be quite constant in shape during a relatively long time period, ..., the drift speed between the two components of the helium and hydrogen ion distributions, respectively, was on the average nearly equal to the local Alfvén speed*'. It should be noted that the dip between the two peaks shown in Figure 4 is not very deep, it may be filled (or partially filled) by other physical processes such as wave-particle interaction with other type waves. However, even if the double peak structure disappears, the asymmetry of the parallel velocity distribution function can still persist.

Figures 5 and 6 are the contour plots of ion velocity distributions in plasma frame at simulation time $\Omega_p t = 2500, 10000, 15000,$ and $20000$ for Case 1 ($\beta \approx 0.1$) and Case 4 ($\beta \approx 0.5$). The top panel is in the $v_x - v_z$ phase plane, and the bottom panel is in the $v_\perp - v_z$ phase plane, where $v_\perp = \sqrt{v_x^2 + v_y^2}$ is the ion velocity perpendicular to the ambient magnetic field. Solid contour lines correspond to fractions of the maximum phase space density 0.8, 0.6, 0.4, and 0.2, and dotted lines correspond to logarithmically spaced fractions 0.1, 0.032, 0.01 and 0.0032. These levels of contour lines are the same as those used in the contour plots for helium by Marsch et al. [1982b]. The distribution features, such as the double streaming peaks and the asymmetry along the ambient magnetic field, are also revealed clearly in these two-dimensional velocity distributions. The double-peak patterns of the distribution in $v_x$-$v_z$ plane are basically similar to those reported in Marsch et al. [1982b]. A common property of the velocity distributions observed by Helios is that the parallel velocity distribution of helium ions along the ambient magnetic field is always more diffusive in the direction outward from the sun than in the direction toward the sun.

One can explain straightly the formation of double streaming peaks from the pickup process described in Section 3. As shown in Figure 3, an ion being picked up means it can be intermittently bounced backward and forward in the Alfvén frame by the LFW field due to magnetic mirror force. In the Alfvén frame, the parallel velocity of this ion along the local magnetic field is approximately zero only near the position where the ion is reflected. At most of the other positions and other times, this ion is streaming along the local magnetic field with a relatively large speed. If we calculate the probability distribution of parallel velocity for this ion in the Alfvén frame, one would expect that there is a dip near zero parallel velocity. Of course, this velocity dip will be located at around the Alfvén speed in the plasma frame. This feature can be seen clearly from Figure 7, in which the top panel illustrates the normalized probability distribution of ion parallel velocity in plasma frame during the simulation time for two trapped ions in Case 1. The time variation of ion position $z$ in the Alfvén frame for these two ions is also shown in the bottom panel, where the red line is for the same ion shown in Figures 2 and 3. The above discussion is for the velocity probability distribution of individual particles at different times. When we take ensemble average for all particles at a given time, the distribution is similar to those shown in Figure 4.

**4.2    Kinetic temperature and differential streaming speed.**

Figure 8 illustrates the time evolution of ion kinetic temperature $T_{kin}$ normalized to $m_{O^{5+}}V_A^2/2$ (top panel), and the ion average parallel velocity $V_{\|O^{5+}}$ normalized to Alfvén speed $V_A$ (bottom panel) for Case 1 ($\beta \approx 0.1$) and Case 4 ($\beta \approx 0.5$). The kinetic temperature is defined by $T_{kin} = \frac{1}{2}m_\alpha \langle (v_z - \langle v_z \rangle)^2 \rangle + \frac{1}{2}m_\alpha [\langle v_x^2 \rangle + \langle v_y^2 \rangle]$, and the average parallel velocity $V_\| = \langle v_z \rangle$. Here, the bracket $\langle \cdot \rangle$ denotes an average over all particles. In both cases, the kinetic temperature and the parallel speed increase continually with simulation time and approach to nearly constant values. The higher the ion initial thermal speed (or plasma beta), the faster the temperature approaches the asymptotic value.

The saturation of the ion kinetic energy and the average velocity can be explained physically based on the pickup process described in the above section. The amplitude of wave satisfies a power-law relation with $B_j^2 \propto \omega_j^{-5/3}$, so most of the wave energy is carried by the LFWs. Moreover, minor ions are mainly picked up or trapped in the wave field of LFWs as demonstrated in the above section. The dispersion of Alfvén waves are small, all wave modes propagate nearly with the same phase speed, namely, the Alfvén speed. Thus, the energy of each ion is almost conserved in the Alfvén frame. For simplicity, we consider the condition of ion energy conservation so that

$$v_\perp^2(t) + \left[v_\|(t) - V_A\right]^2 = v_\perp^2(0) + \left[v_\|(0) - V_A\right]^2, \tag{4}$$

Where $v_\perp$ and $v_\parallel$ are the velocity components perpendicular and parallel to the ambient magnetic field $\mathbf{B}_0$. The motion of ions on a spherical surface in velocity space, defined by Eq. (4), is caused by pitch-angle scattering from Alfvén waves. With increasing time, as shown in the top panel of Figure 1, more and more ions are pitch-angle scattered from the left side of the spherical shell to the right side, or vice versa. Gradually, a spherical shell-like distribution is formed. When a full spherical shell-like distribution is formed, the pickup process is completed. Because the center of the spherical shell is moving with the Alfvén speed in the plasma frame, this also implies that minor ions gain a bulk parallel velocity roughly equal to the Alfvén speed. In other words, minor ions flow faster than protons along the ambient magnetic field $\mathbf{B}_0$ at a relative speed roughly equal to the Alfvén speed.

Assuming the ion initial velocity distribution is a Maxwellian distribution, it is easy to demonstrate by taking ensemble average of Eq. (4) that time variation of the normalized minor ion kinetic temperature and average parallel speed approximately satisfies the following equation

$$T_{kin}(t) = 1 + T_{kin}(0) - \left[1 - V_\parallel(t)\right]^2, \tag{5}$$

where the kinetic temperature is normalized to $m_{O^{5+}} V_A^2 / 2$, and average speed is normalized to the Alfvén speed. The simulation results for $T_{kin}(t)$ versus $V_\parallel(t)$ for Cases $\beta \approx 0.1$ and 0.5 are plotted in Figure 9.

## 4.3 Temperature anisotropy versus differential streaming speed

The bottom panel of Figure 10 shows the variation of kinetic temperature anisotropy $T_{\perp O^{5+}} / T_{\parallel O^{5+}}$ versus the average parallel velocity $V_{\parallel O^{5+}}$ at different times for simulation Case 1 ($\beta \approx 0.1$) and Case 4 ($\beta \approx 0.5$). Because Alfvén waves are mainly carried by protons, the average parallel velocity $V_{\parallel O^{5+}}$ in Figure 10 represents the differential streaming speed between minor ions and protons in observations. The temperature anisotropy firstly increases with the increase of average velocity to a maximum value about 4.5 (in case $\beta \approx 0.1$) or 2.0 (in case $\beta \approx 0.5$) at $V_{\|O^{5+}} \approx 0.2V_A$, and then it decreases with further increase of average velocity. When the average velocity is greater than $0.4V_A$, the anisotropy reaches a constant value slightly less than one ($T_{\perp O^{5+}} < T_{\|O^{5+}}$). To our knowledge, there is no *in situ* observation on the temperature anisotropy for minor ions such as $O^{5+}$ in solar wind. However, the observational relationship between the temperature anisotropy of alpha particles and the differential streaming speed between alpha particles and protons has been reported in a number of papers [e. g., Bourouaine et al. 2011a; Kasper et al. 2008; Gary et al. 2005]. This trend in our simulation is generally consistent with observational trend shown in Figure 4 of Kasper et al. [2008], and in Figure 6b and Figure 7b of Gary et al. [2005]. Their results show that large values of the alpha anisotropy are clustered near the region where $\Delta V_{\alpha p} \leq (0.1 \sim 0.2)V_A$. When the differential streaming speed is greater than $0.2V_A \sim 0.3V_A$, the alpha anisotropy does not change significantly with the increase of differential speed and is generally less than unity. For convenience of the reader, Figure 6b in Gary et al. [2005] is re-plotted as the top panel in Figure 10. We reiterate that both helium and other minor ions are observed to be hotter than protons and flowing faster than protons with a relative speed less or equal to the Alfvén speed in solar wind.

5. **Pickup and Kinetic Properties of Minor Ions in the Solar Corona**

The plasma beta value and the normalized power of Alfvén-cyclotron waves are very low in the solar corona, comparing with the values in solar wind. Case 5 and Case 6 in table 1 are for the condition in the solar corona. The initial thermal speed of ions is $v_{TO^{5+}} = 0.05V_A$ in both cases, corresponding to a plasma beta value $\beta = 2.5 \times 10^{-3}$, while the normalized wave energy density $\delta B^2 / B_0^2$ equals 0.01 and 0.02 for Cases 5 and 6, respectively.

Figure 11 illustrates the time variations of ion kinetic temperature, temperature anisotropy and average parallel velocity for simulation Case 5 (red line) and Case 6 (black line). It is found that $O^{5+}$ ion can be heated quickly to a high kinetic temperature $T_{kin}/(m_{O^{5+}}V_A^2/2) \approx 0.3$ within a few hundred gyro-periods of proton in both cases. At the same time, $O^{5+}$ ions are accelerated to an average parallel velocity roughly equal to a value $0.17V_A$. After that time, there is only slight heating and accelerating. Clearly, $O^{5+}$ ion can be only partially picked up by the Alfvén-cyclotron waves due to the low density of wave energy in the corona. When the ion is stochastically heated, its temperature anisotropy is generally greater than 10 with a maximum anisotropy about 20.

If we consider the Alfvén speed $V_A \approx 1000 - 2000$ km s$^{-1}$ in the upper solar corona, the out flow speed of $O^{5+}$ ions would be faster than that of protons about $0.2V_A \approx 200 - 400$ km s$^{-1}$ based on the above simulations. At the same time, the kinetic temperature $T_{kin} \approx T_{kin,\perp} \approx 0.3 \times m_{O^{5+}}V_A^2/2$ implies that the $O^{5+}$ ions could have a perpendicular thermal speed $v_{th\perp}(O^{5+}) \approx 300 - 600$ km s$^{-1}$. These results are generally consistent with the SOHO observations (Kohl et al. 1997, 1998; Cranmer et al. 1999a, 1999b). It is found that the $O^{5+}$ ions have a perpendicular temperature $T_\perp \sim 2 \times 10^8$ K corresponding to a thermal speed $\sim 450$ km s$^{-1}$, and the different outflow speed between $O^{5+}$ ions and protons is about $200 - 300$ km s$^{-1}$ in the corona at heliocentric distance of about 2 to 4 solar radii. The high temperature anisotropy in the simulation is also consistent with observations.

## 6. Compound Effect of Alfvén Waves and Ion Cyclotron Waves

In this section, a number of other simulation cases with different wave energy densities and plasma $\beta$ values are performed. Figure 12 shows the variation of the final values of ion kinetic temperature and average parallel velocity with the normalized wave energy density $\delta B^2/B_0^2$ at the end of the simulation time $\Omega_p t = 20000$ for different simulation cases, where $\delta B^2 / B_0^2$ is the value for the whole wave spectrum in the frequency range $[0.01, 0.4]\Omega_p$. Similar to the discussion in section 3, three Cases of A, H and L are studied for each value of $\delta B^2 / B_0^2$. In Case A, the whole wave spectrum is included, namely, all the discretized wave modes of number $j = 1, 2, ..., 51$ are used. In Cases H and L, we only use the wave modes of number $j = 36, 37, ..., 51$ with frequency range $(0.12, 0.4]\Omega_p$ (mainly HFW ion-cyclotron waves), and the wave modes of number $j = 1, 2, ..., 35$ with frequency range $[0.01, 0.12]\Omega_p$ (LFW Alfvén waves), respectively. The black line, red line and green line in Figure 12 show the results for Cases A, H and L, respectively. The blue line is the sum of values obtained in Case H and Case L separately. The results for plasma $\beta = 0.1$ and 0.01 are illustrated by the solid lines and dash lines.

It is found from Figure 12 that both ion kinetic temperature and average velocity increase monotonically with the increase of wave energy density ratio $\delta B^2 / B_0^2$. However, their values increase very slowly after $\delta B^2 / B_0^2$ is greater than 0.1. This agrees with the result of Case 1 in Section 3 that a spectrum of Alfvén-cyclotron waves with normalized wave energy density equal to 0.1 is high enough to almost fully pick up the minor ions within the simulation time. When minor ions are fully picked up, the ion kinetic temperature and average parallel speed approach values dictated by the Alfvén speed that are independent of the wave amplitude. The increment of ion kinetic temperature and average velocity does not depend significantly on the plasma beta value.

A very interesting result is that there is a compound effect of Alfvén waves and ion-cyclotron waves on the heating and acceleration of minor ions via the pickup process. One can see from Figure 12 that the final kinetic temperature and average parallel speed obtained from simulation Case A are generally larger than the sum of values obtained separately from Case H and Case L, when the wave energy density ratio is less than 0.2. In other words, the heating and acceleration efficiency in the case that the Alfvén-cyclotron waves are considered as a whole spectrum, is larger than the efficiency in the case that Alfvén waves and ion-cyclotron waves are considered separately and then their efficiencies are added up. This compound effect is mainly manifested in the heating and parallel acceleration of ions (the bottom panels in Figure 12). Heating of ions in the perpendicular direction is mainly produced by ion-cyclotron waves.

Based on the above simulation results, we consider that there are two stages for the heating and acceleration of minor ions by Alfvén-cyclotron waves in the solar corona and solar wind. Firstly, minor ions are mainly heated in the perpendicular direction by high frequency ion-cyclotron waves due to resonant wave-particle interactions. A temperature anisotropy maximum is obtained at this stage. Secondly, minor ions are farther pitch-angle scattered in both the perpendicular and the parallel direction by the low frequency Alfvén waves if the energy density of Alfvén waves is relatively high. At this stage, both the heating and the acceleration of ions are greatly enhanced especially in parallel direction, with a decrease of the temperature anisotropy. If the normalized wave energy density in the solar wind is high enough, a full sphere-shell-like distribution would be formed as in Case 1 and Case 4. However, this is not easy to happen for minor ions in the solar corona due to the low level of wave energy density ratio ($\delta B^2 / B_0^2 < 0.02$). In addition, we would like to reiterate that ion-cyclotron waves also play an important role at the second stage by randomizing the ion orbit through resonant wave-particle interaction.

7. **Summary**

In this paper, we present a scenario to explain the preferential heating of minor ions and the differential streaming speed between minor ions and protons observed in the solar coronal and in the solar wind. The basic idea is that minor ions may be picked up (or partially picked up) by Alfvén-cyclotron waves, which pervade intrinsically in the solar corona and interplanetary space. The main results are summarized as follows:

1. It is demonstrated by test particle simulations that minor ions can be nearly fully picked up by Alfvén-cyclotron waves in the solar wind, based on the observed wave energy density with $\delta B^2 / B_0^2 \geq 0.1$ by Helios at 0.7 AU. It is found that both high frequency ion-cyclotron waves and low frequency Alfvén waves may play key roles for the pickup of minor ions in the solar wind. A minor ion can obtain a high magnetic moment through resonant wave-particle interactions with HFWs. The ion with large magnetic moment can be intermittently bounced backward and forward or trapped by magnetic mirror-like field structures formed by the large amplitude LFWs. When an ion is trapped in the wave field, the ion is picked up by these waves.

2. It is found that after being picked up by Alfvén-cyclotron waves in the solar wind, the parallel velocity distribution of minor ions is asymmetric and has double streaming peaks with a dip at Alfvén speed along the ambient magnetic field. The distance between two peaks is approximately equal to the Alfvén speed. The formation of double streaming peaks with a dip can be understood based on the pickup process. The reason is that the ion, most of time, is streaming backward and forward along the magnetic field, and only occasionally reflected near the magnetic mirror point to have a small parallel velocity in the Alfvén frame.

3. During the pickup process, minor ions are stochastically heated and accelerated by Alfvén-cyclotron waves. Minor ions can flow faster than protons with a relative speed less or equal to the Alfvén speed. When minor ions are only partially picked up by waves，a partially spherical shell-like velocity distribution is formed, and minor ions would have a highly anisotropic temperature. With an increasing number of ions being fully picked up by waves, a fully spherical shell-like velocity distribution can be formed. At this time, the ion kinetic temperature becomes approximately isotropic with the parallel temperature slightly greater than the perpendicular temperature.

4. Both alpha particles and other minor ions are hotter than and flow faster than protons in solar wind from observations. There is no *in situ* observation on the velocity distribution and temperature anisotropy of minor ions except helium ions. If we assume both helium and other minor ions are picked up by the intrinsic Alfvén-cyclotron waves in the solar wind, the kinetic properties of ions from simulations are generally consistent with *in situ* features observed in the solar wind, such as the asymmetry and double-peak structures of the ion parallel velocity distribution with respect to the ambient magnetic field, and the variation of the temperature anisotropy with the differential streaming speed between alphas and protons [e.g. Marsch et al. 1982b; Gary et al., 2005; Kasper et al. 2008].

5. Minor ions can only be partially picked up by Alfvén-cyclotron waves in the solar corona, because the plasma beta value and the normalized power of Alfvén-cyclotron waves are very low in the solar corona compared with their values in the solar wind. The simulation results are also consistent with the remotely observed kinetic features of minor ions such as $O^{5+}$ in the upper solar corona [e.g. Kohl et al. 1997, 1998; Li et al., 1998; Cranmer et al., 1999a, 1999b; Kohl et al., 2006].

6. There exists a compound effect of Alfvén waves and ion-cyclotron waves on the heating and acceleration of minor ions in the solar corona and solar wind. Minor ions can be strongly heated in the perpendicular direction by ion-cyclotron waves due to resonant wave-particle interactions. Then they are farther pitch-angle scattered by the low frequency Alfvén waves. As a result, both heating and parallel acceleration of ions are greatly enhanced.

Historically, the pickup of ions by Alfvén waves in solar wind has been studied extensively since the early 1970s [e.g. Wu & Davidson, 1972; Winske et al., 1985; Bogdan et al., 1991; Zank, 1999, and references therein]. There is a major difference between the pickup process discussed in this paper and that in the literature. Most of previous discussions are focused on the pickup of ionized neutrals due to charge exchange, electron impact, or photoionization. These newly born ions initially have a large streaming speed in the solar wind frame (or plasma frame) that is much greater than the local Alfvén speed. The question to be asked is whether these newborn ions would co-move or be "picked up" by the solar wind. In the present study, minor ions are initially with a zero bulk velocity in the plasma frame, in other words, they are initially co-moving with the background plasma or protons. It is demonstrated that they will be picked up by the intrinsic Alfvén waves observed in the solar wind, and as a result, they will have a different streaming speed with respect to the background protons after being picked up and moving with waves.

In addition, we would like to give some further discussion on the approximations made for simplicity and distilling the essential physics in this study. Firstly, all wave modes are assumed to be outward propagating with respect to the Sun in the solar wind frame. It is found from observations that there may be inward Alfvénic fluctuations in solar wind [e.g. Bavassano, Pietropaolo and Bruno, 2001]. We believe that the scenario proposed in this paper is still applicable even if there is counter-propagating waves, because the power of outward Alfvénic waves is dominant in the solar wind. Detail analysis of counter-propagating wave effects can be considered in the future, which is beyond the scope of the present paper. Secondly, the magnetic field lines expand with altitude in the corona. In the presence of diverging field lines, the temperature anisotropy ($T_\perp > T_\parallel$) provides the ions with an upward lifting force (Lee & Wu, 2000). This upward force will convert the perpendicular thermal energy to parallel flow energy and help push the perpendicularly-heated minor ions outward.

The Alfvén speed decreases with increasing heliospheric distance from a few thousand km/s in the upper corona at several solar radii to tens km/s at 1 AU and beyond in the interplanetary space. The different streaming speed between minor ions and protons is about 200-300 km/s in the upper corona, which is much larger than local the Alfvén speed in the interplanetary space at 1AU. Since the proton speed remains roughly constant in the interplanetary space, it remains a puzzle how minor ions are being decelerated in the interplanetary space. Based on the scenario proposed in this paper, if minor ions are picked up by or trapped in the low frequency Alfvén wave field, they will be decelerated accordingly with the deceasing of Alfvén speed.

Finally, a common feature of proton velocity distribution functions observed in fast solar wind is that in addition to the anisotropic core distribution, there is a second proton beam component travelling at about 1.5 times of the local Alfvén speed [Marsch et al., 1982c]. The physical process discussed in this paper may also help us to understand the formation of the proton beams if a few percent of protons are trapped in the low frequency Alfvén wave field.

**Acknowledgments** The research at USTC was supported in part by the National Nature Science Foundation under grants 41174123, 40931053 and 41121003, and in part by the Chinese Academy of Sciences under grants KZCX2-YW-QN512 and KZZD-EW-01. The research at Academia Sinica was supported by the National Science Council (NSC-101-2628-M-001-007-MY3) in Taiwan. C.B.W. acknowledges the hospital invitation of Prof. J. H. Shue for his visiting in NCU in 2012.

**Table 1:** Parameters for different simulation cases.

| | $\varepsilon_0 = \dfrac{\delta B^2}{B_0^2}$ | $[\omega_1, \omega_{N=51}](\Omega_p)$ | $\beta = \dfrac{v_T^2}{V_A^2}$ | Range of wave number j used in simulation |
|---|---|---|---|---|
| Case 1 | 0.1 | [0.01, 0.4] | 0.1 | j = 1~51 (all waves) |
| Case 2 | 0.1 | [0.01, 0.4] | 0.1 | j = 42~51 (only high frequency waves) |
| Case 3 | 0.1 | [0.01, 0.4] | 0.1 | j = 1~35 (only low frequency waves) |
| Case 4 | 0.1 | [0.01, 0.4] | 0.5 | j = 1~51 (all waves) |
| Case 5 | 0.01 | [0.01, 0.4] | $2.5 \times 10^{-3}$ | j = 1~51 (all waves) |
| Case 6 | 0.02 | [0.01, 0.4] | $2.5 \times 10^{-3}$ | j = 1~51 (all waves) |

*Cases 1-4 are for the parameters common in the solar wind; Cases 5-6 are for parameters in the solar corona.

**Figure Captions**

**Figure 1** The velocity scatter-plot of alpha particles in the $v_z$-$v_\perp$ phase plane in the plasma frame at different simulation time, where $v_z$ and $v_\perp = \sqrt{v_x^2 + v_y^2}$ are velocities in the direction parallel and perpendicular to the ambient magnetic field. The top, medium and bottom panels are results for Case 1, Case 2 and Case 3, respectively.

**Figure 2** Variation of magnetic moment and position $z$ of an ion in the Alfvén frame with simulation time. The red, green and blue lines represent the results of Case 1, Case 2 and Case 3, respectively. Each ion has the same initial velocity and position in all three simulation cases. The red short vertical bars indicate times when the ion is reflected by large amplitude wave field due to magnetic mirror force for Case 1.

**Figure 3** The red dash lines show the projected trajectories of an ion along the ambient magnetic field in the Alfvén frame at three selected time intervals in Case 1. The solid curves represent the strength of wave magnetic field experienced by this ion at different position. This ion is the same ion shown in Figure 2. The reflecting times are also indicated by the red short vertical bars in Figure 2.

**Figure 4** The normalized ion velocity distribution in the plasma frame at simulation time $\Omega_p t = 0$ (black solid line), 2500 (red dot line), 10000 (green dash line), 20000 (blue dash-dot line). The top and bottom panel is for Case 1 ($\beta = 0.1$) and Case 4 ($\beta = 0.5$), respectively.

**Figure 5** Contour plots of ion velocity distribution in the plasma frame at simulation time $\Omega_p t = 2500$, 10000, 15000, 20000 for Case 1 ($\beta = 0.1$). The top panel is in the $v_x$-$v_z$ phase

plane, while the bottom panel is in the $v_z$-$v_\perp$ phase plane, **where** $v_\perp = \sqrt{v_x^2 + v_y^2}$. Contour lines correspond to fractions of the maximum phase density 0.8, 0.6, 0.4, and 0.2 (solid lines) and to logarithmically spaced fractions 0.1, 0.032, 0.01 and 0.0032 (dash lines).

**Figure 6** Same as Figure 5, but for simulation Case 4 ($\beta = 0.5$).

**Figure 7** The top panel is the normalized probability distribution of ion parallel velocity in the plasma frame during the simulation time period for two trapped ions in Case 1. One of the two trapped ions is the same ion shown in Figures 2 and 3 and the results are shown by red lines. The bottom panel is the time variation of ion position $z$ in the Alfvén frame for these two ions.

**Figure 8** Time evolution of ion kinetic temperature (top panel) and ion average parallel velocity (bottom panel) in the plasma frame for simulation Case 1 ($\beta = 0.1$) and Case 4 ($\beta = 0.5$).

**Figure 9** Variation of ion kinetic temperature with the ion average parallel velocity in the plasma frame at different simulation times (time increases from left to right) for Case 1 ($\beta = 0.1$) and Case 4 ($\beta = 0.5$).

**Figure 10** Variation of temperature anisotropy (bottom panel) with the ion average parallel velocity in the plasma frame at different simulation times (time increases from left to right) for Case 1 ($\beta = 0.1$) and Case 4 ($\beta = 0.5$). The top panel shows the variation of temperature anisotropy versus the differential streaming speed $V_{\alpha p}$ between alpha particles and protons in the solar wind observed by ACE satellite [from Gary et al., 2005], where dash line represents least squares fits to the data. The colors represent the number of observations which lie within each pixel. The scales are arbitrarily chosen to convey the data trends, and range from black (no observation) through purple, blue, green, yellow, orange, and red (most observations).

**Figure 11** Time evolution of ion kinetic temperature (top panel), kinetic temperature anisotropy (medium panel) and ion average parallel velocity (bottom panel) in the plasma frame for Case 5 (red line) with $\delta B^2 / B_0^2 = 0.01$ and $\beta = 0.0025$ and Case 6 (black line) with $\delta B^2 / B_0^2 = 0.02$ and $\beta = 0.0025$.

**Figure 12** Variation of the ion kinetic temperature and average parallel speed with the normalized wave energy density $\delta B^2 / B_0^2$ at the end of the simulation time $\Omega_p t = 20000$, where $\delta B^2 / B_0^2$ is the value of the whole wave spectrum in the frequency range $[0.01, 0.4]\Omega_p$. The black lines, red lines and green lines in figure 11 represent the results for Cases A, H and L, respectively. The blue line is the sum of values obtained in Case H and Case L separately. The results for plasma $\beta = 0.1$ and 0.01 are illustrated by the solid lines and dash lines.

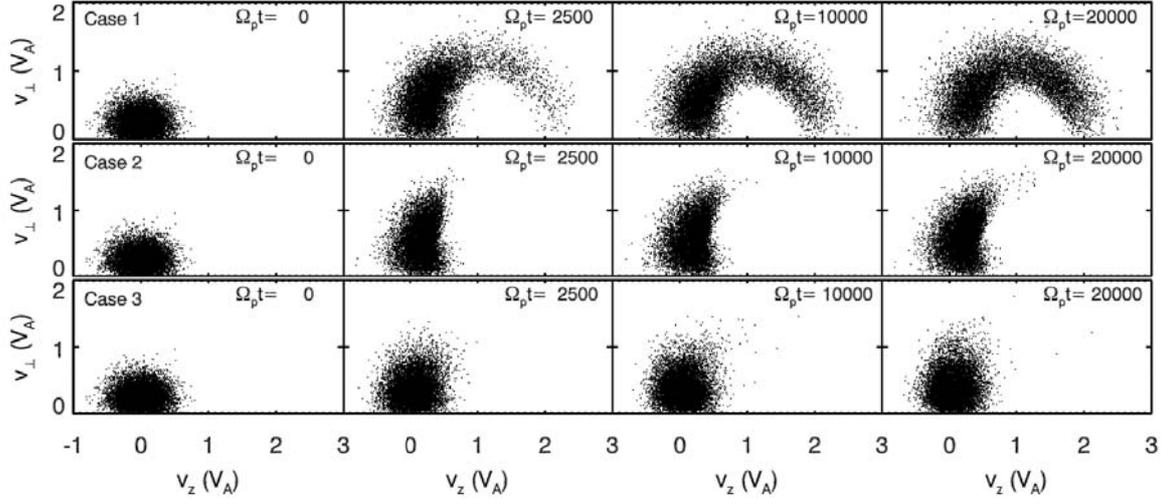

**Figure 1** The velocity scatter-plot of alpha particles in the $v_z$-$v_\perp$ phase plane in the plasma frame at different simulation time, where $v_z$ and $v_\perp = \sqrt{v_x^2 + v_y^2}$ are velocities in the direction parallel and perpendicular to the ambient magnetic field. The top, medium and bottom panels are results for Case 1, Case 2 and Case 3, respectively.

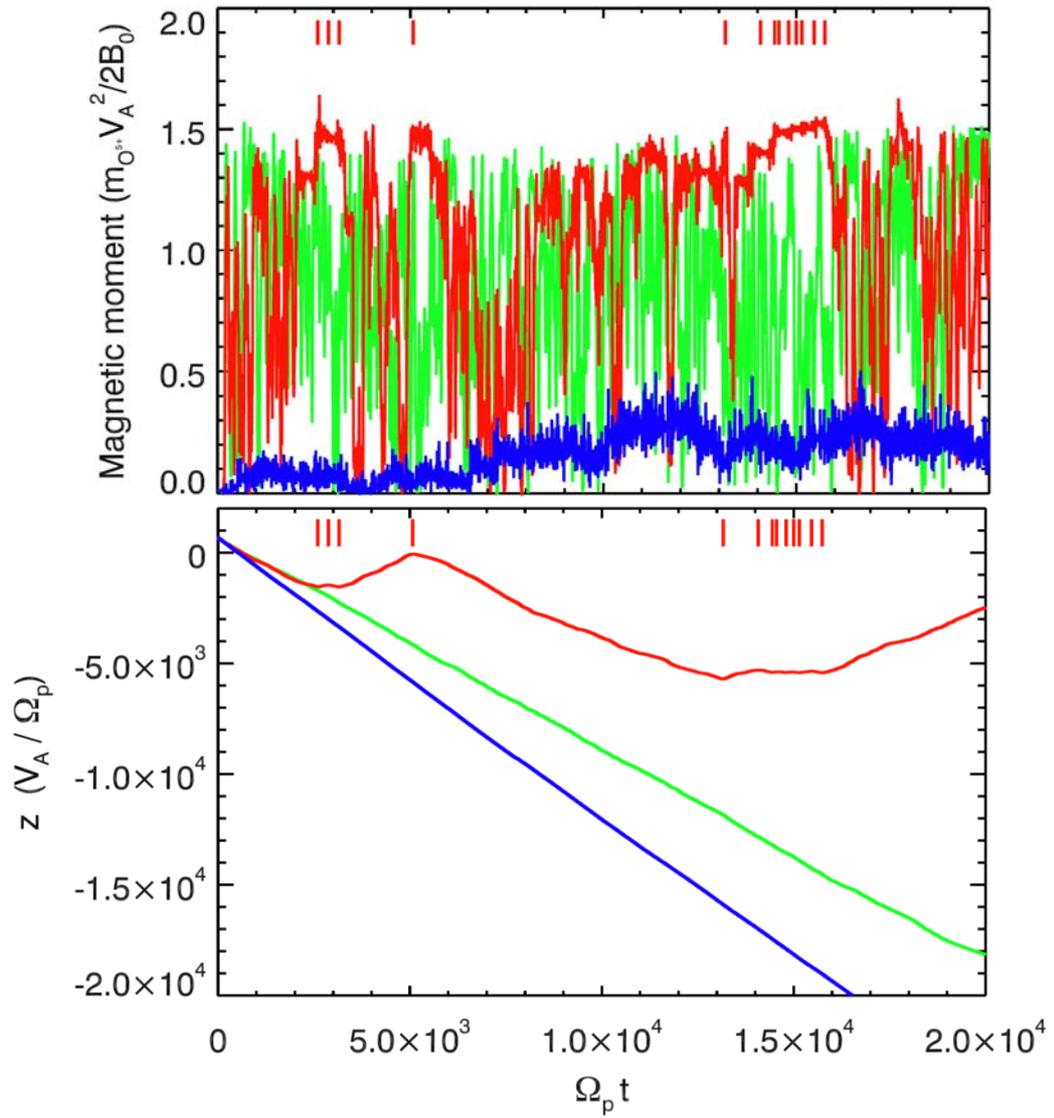

**Figure 2** Variation of magnetic moment and position *z* of an ion in the Alfvén frame with simulation time. The red, green and blue lines represent the results of Case 1, Case 2 and Case 3, respectively. Each ion has the same initial velocity and position in all three simulation cases. The red short vertical bars indicate times when the ion is reflected by large amplitude wave field due to magnetic mirror force for Case 1.

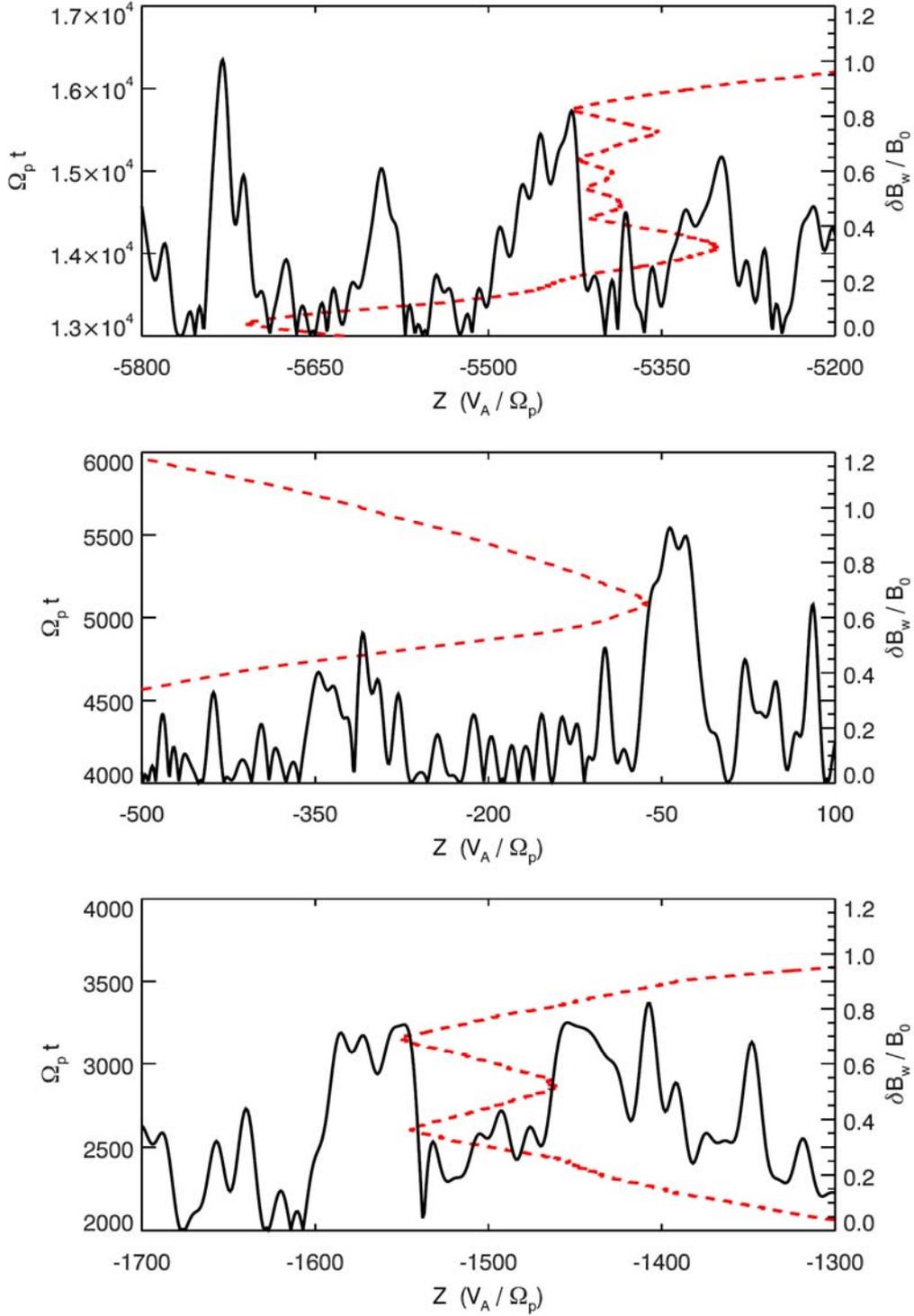

**Figure 3** The red dash lines show the projected trajectories of an ion along the ambient magnetic field in the Alfvén frame at three selected time intervals in Case 1. The solid curves represent the strength of wave magnetic field experienced by this ion at different position. This ion is the same ion shown in Figure 2. The reflecting times are also indicated by the red short vertical bars in Figure 2.

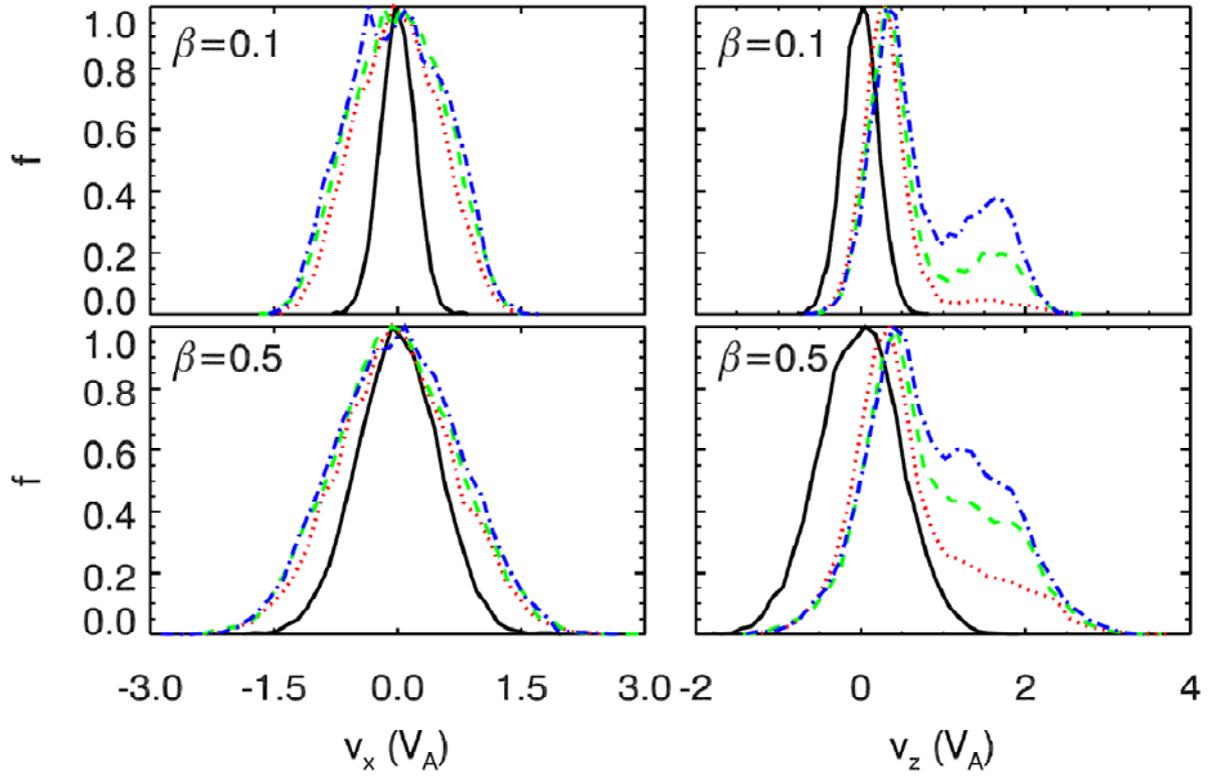

**Figure 4** The normalized ion velocity distribution in the plasma frame at simulation time $\Omega_p t = 0$ (black solid line), 2500 (red dot line), 10000 (green dash line), 20000 (blue dash-dot line). The top and bottom panel is for Case 1 ($\beta = 0.1$) and Case 4 ($\beta = 0.5$), respectively.

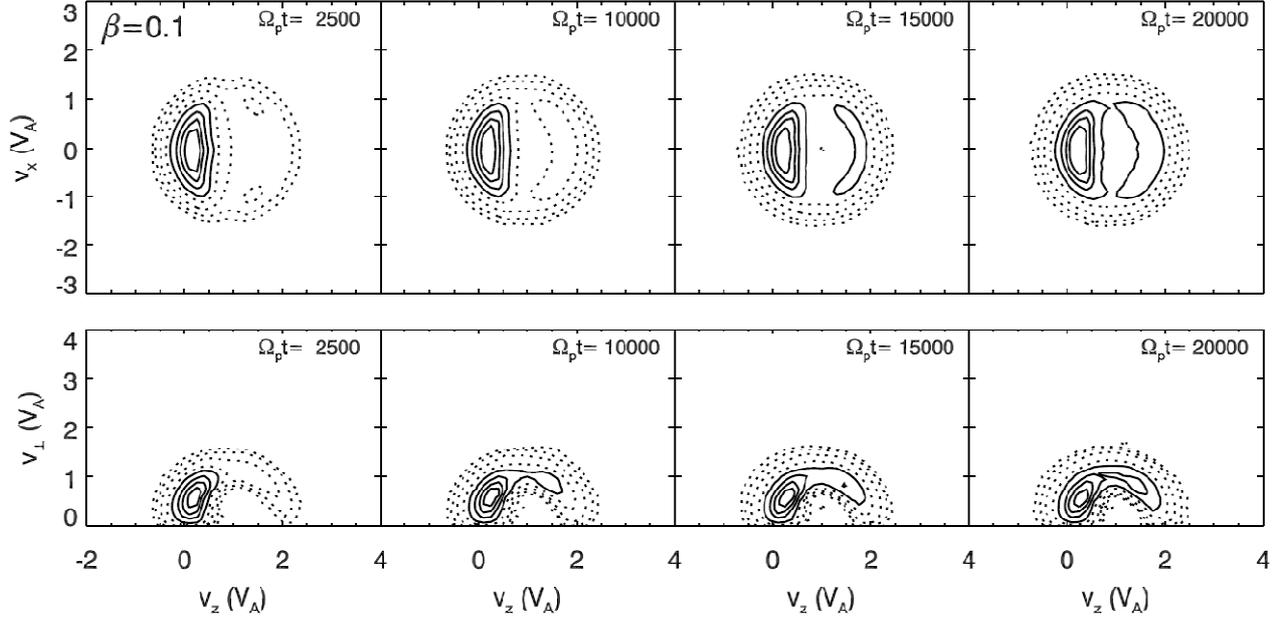

**Figure 5** Contour plots of ion velocity distribution in the plasma frame at simulation time $\Omega_p t = 2500,\ 10000,\ 15000,\ 20000$ for Case 1 ($\beta = 0.1$). The top panel is in the $v_x$-$v_z$ phase plane, while the bottom panel is in the $v_z$-$v_\perp$ phase plane, where $v_\perp = \sqrt{v_x^2 + v_y^2}$. Contour lines correspond to fractions of the maximum phase density 0.8, 0.6, 0.4, and 0.2 (solid lines) and to logarithmically spaced fractions 0.1, 0.032, 0.01 and 0.0032 (dash lines).

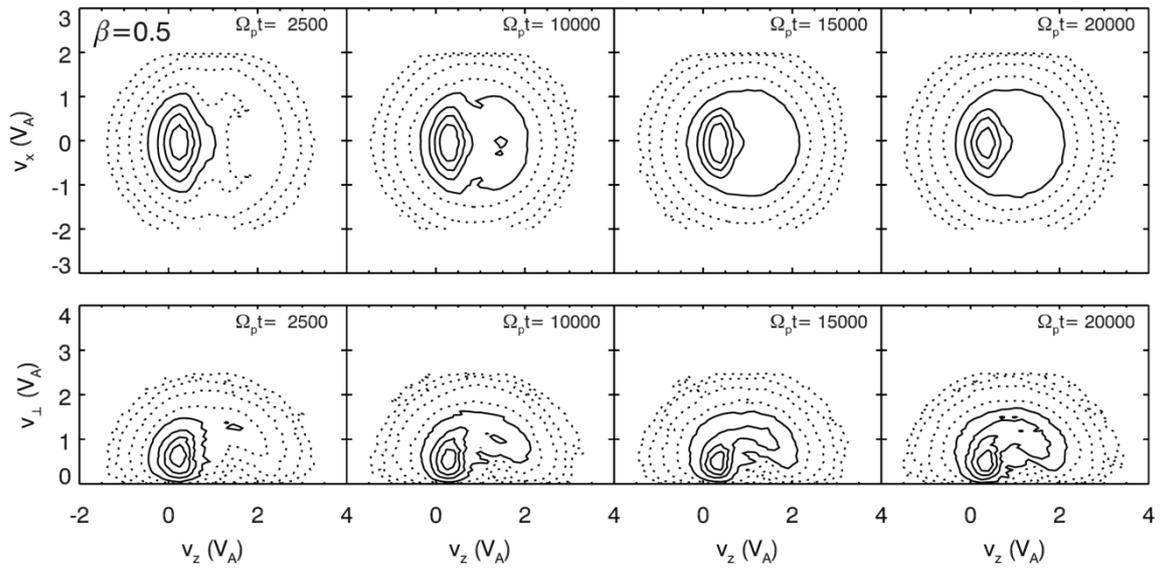

**Figure 6** Same as Figure 5, but for simulation Case 4 ($\beta = 0.5$).

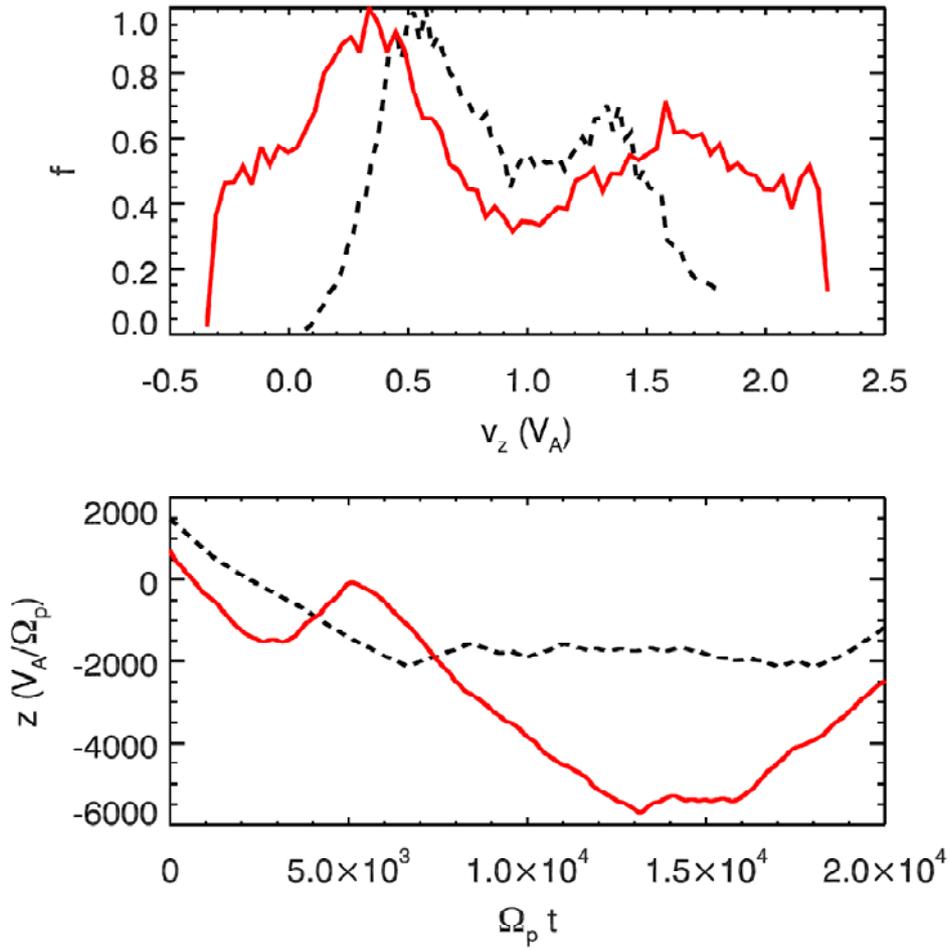

**Figure 7** The top panel is the normalized probability distribution of ion parallel velocity in the plasma frame during the simulation time period for two trapped ions in Case 1. One of the two trapped ions is the same ion shown in Figures 2 and 3 and the results are shown by red lines. The bottom panel is the time variation of ion position $z$ in the Alfvén frame for these two ions.

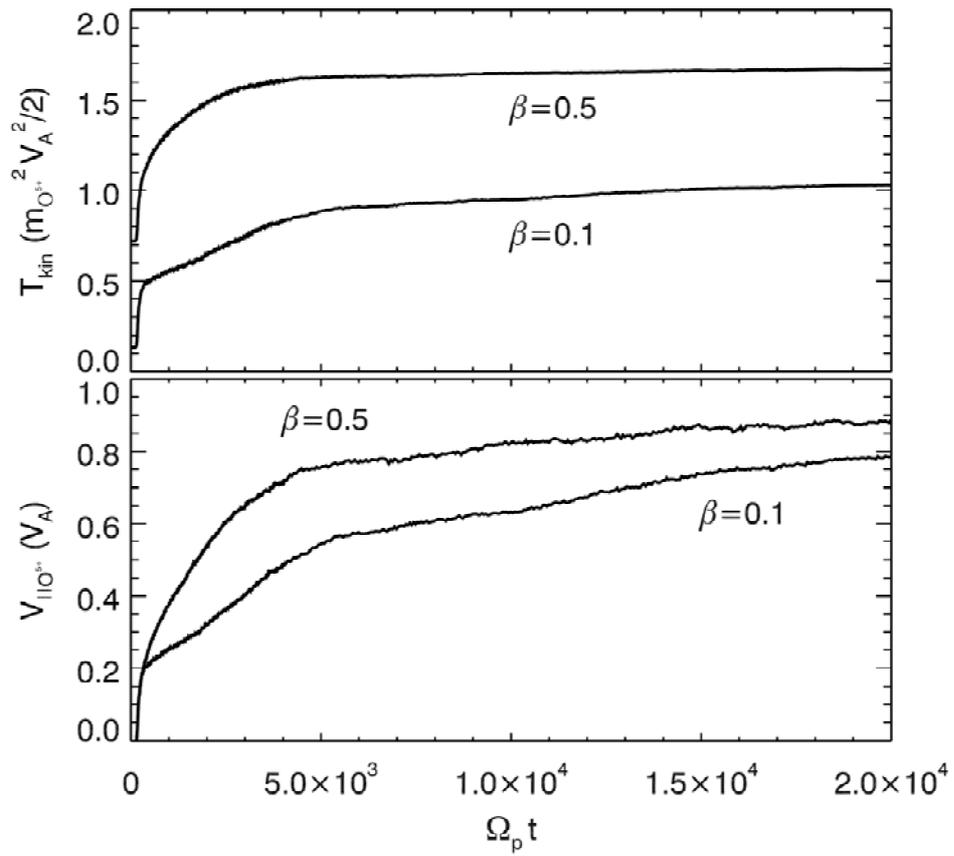

**Figure 8** Time evolution of ion kinetic temperature (top panel) and ion average parallel velocity (bottom panel) in the plasma frame for simulation Case 1 ($\beta = 0.1$) and Case 4 ($\beta = 0.5$).

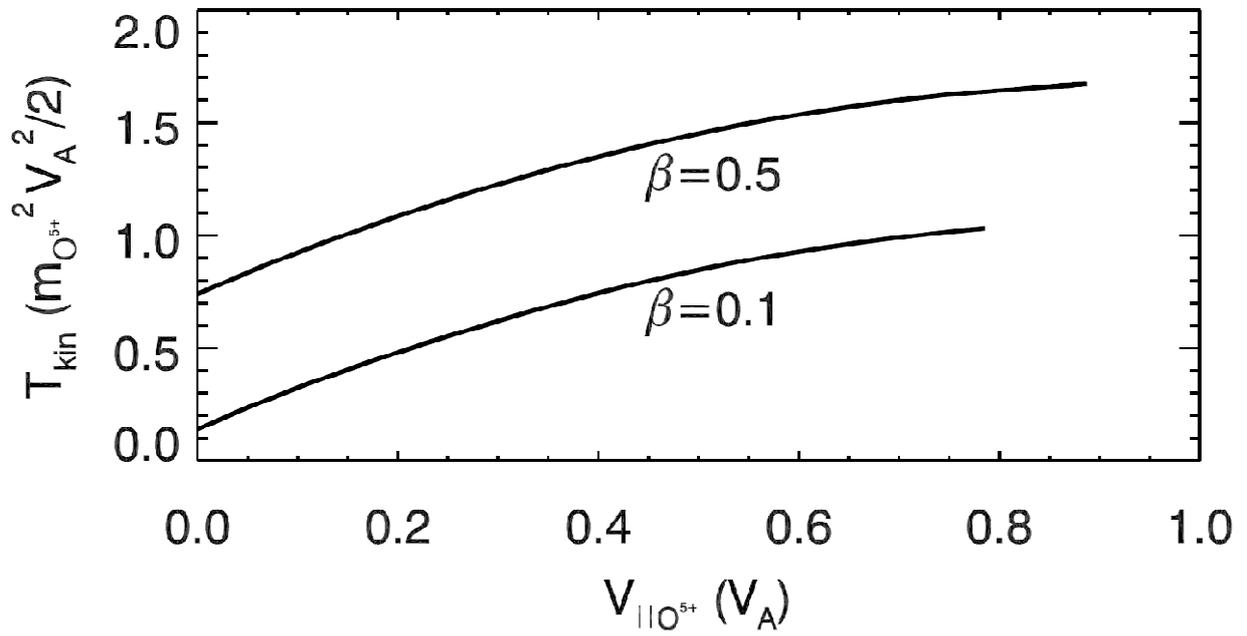

**Figure 9** Variation of kinetic temperature with the ion average parallel velocity in the plasma frame at different simulation times (time increases from left to right) for Case 1 ($\beta = 0.1$) and Case 4 ($\beta = 0.5$).

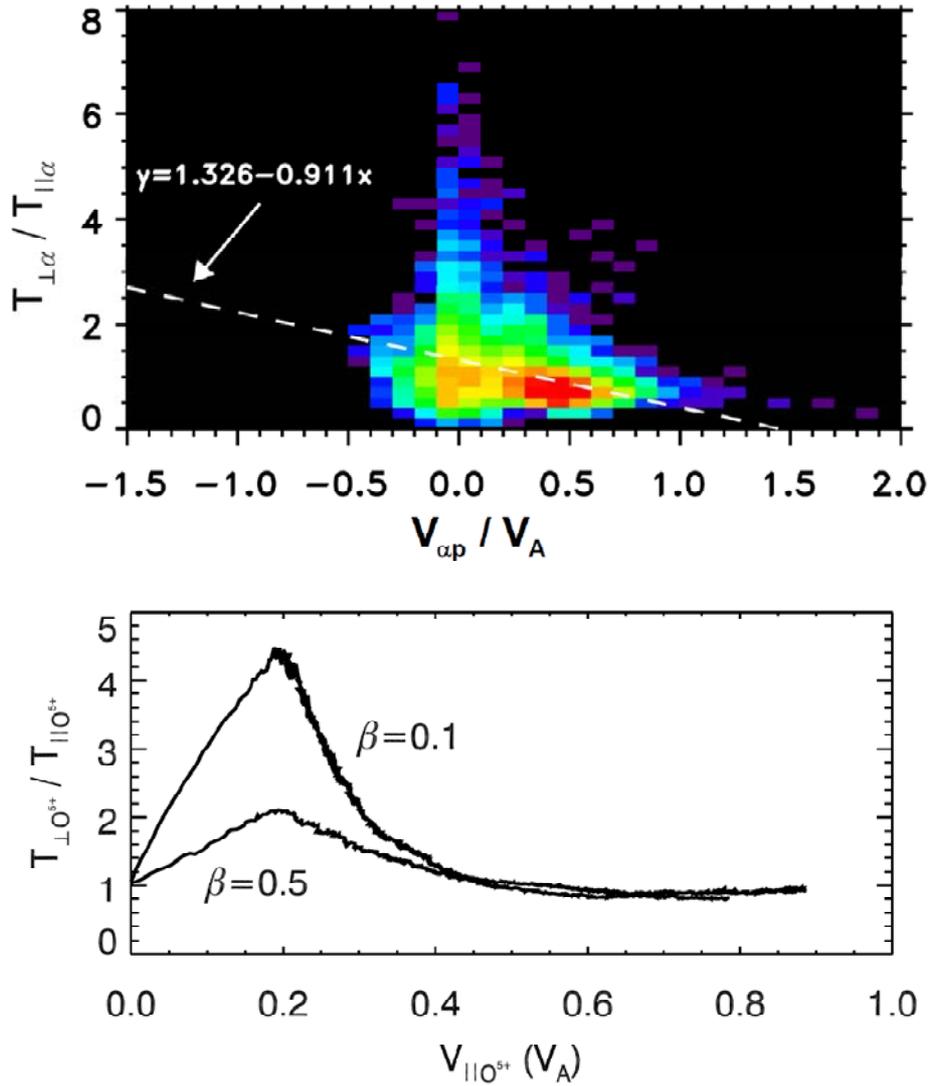

**Figure 10** Variation of temperature anisotropy (bottom panel) with the ion average parallel velocity in the plasma frame at different simulation times (time increases from left to right) for Case 1 ($\beta = 0.1$) and Case 4 ($\beta = 0.5$). The top panel shows the variation of temperature anisotropy versus the differential streaming speed $V_{\alpha p}$ between alpha particles and protons in the solar wind observed by ACE satellite [from Gary et al., 2005], where the dash line represents least squares fits to the data. The colors represent the number of observations which lie within each pixel. The scales are arbitrarily chosen to convey the data trends, and range from black (no observation) through purple, blue, green, yellow, orange, and red (most observations).

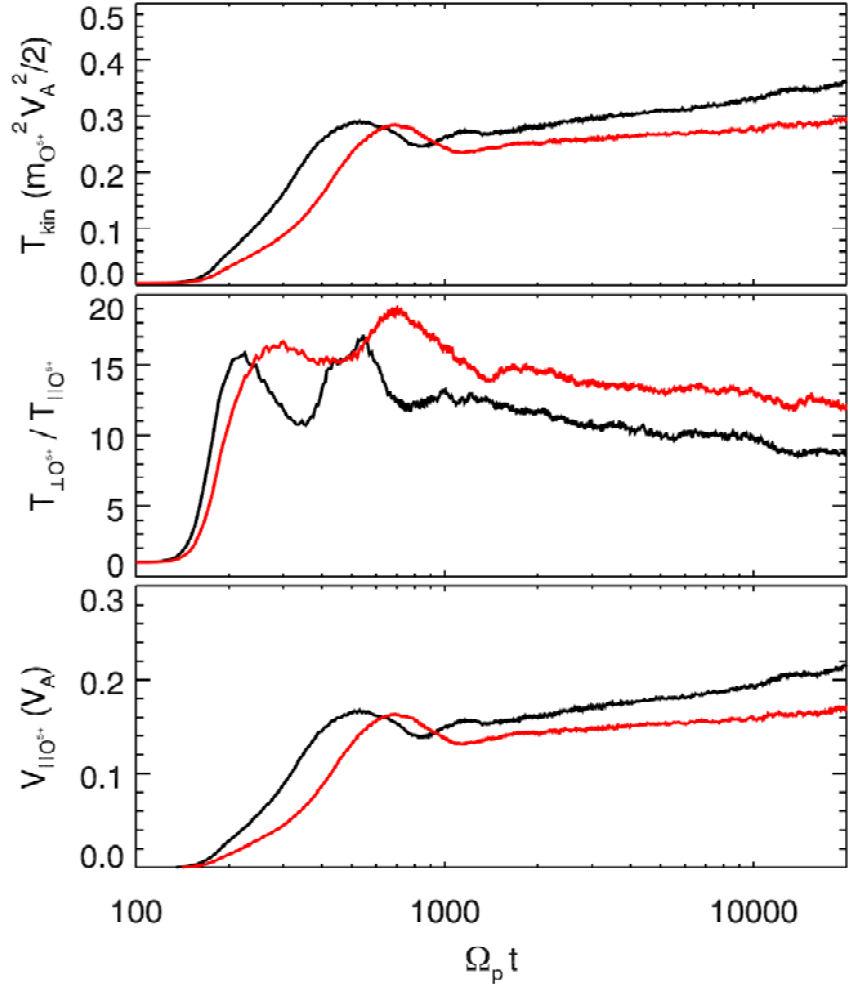

**Figure 11** Time evolution of ion kinetic temperature (top panel), kinetic temperature anisotropy (medium panel) and ion average parallel velocity (bottom panel) in the plasma frame for Case 5 (red line) with $\delta B^2 / B_0^2 = 0.01$ and $\beta = 0.0025$ and Case 6 (black line) with $\delta B^2 / B_0^2 = 0.02$ and $\beta = 0.0025$.

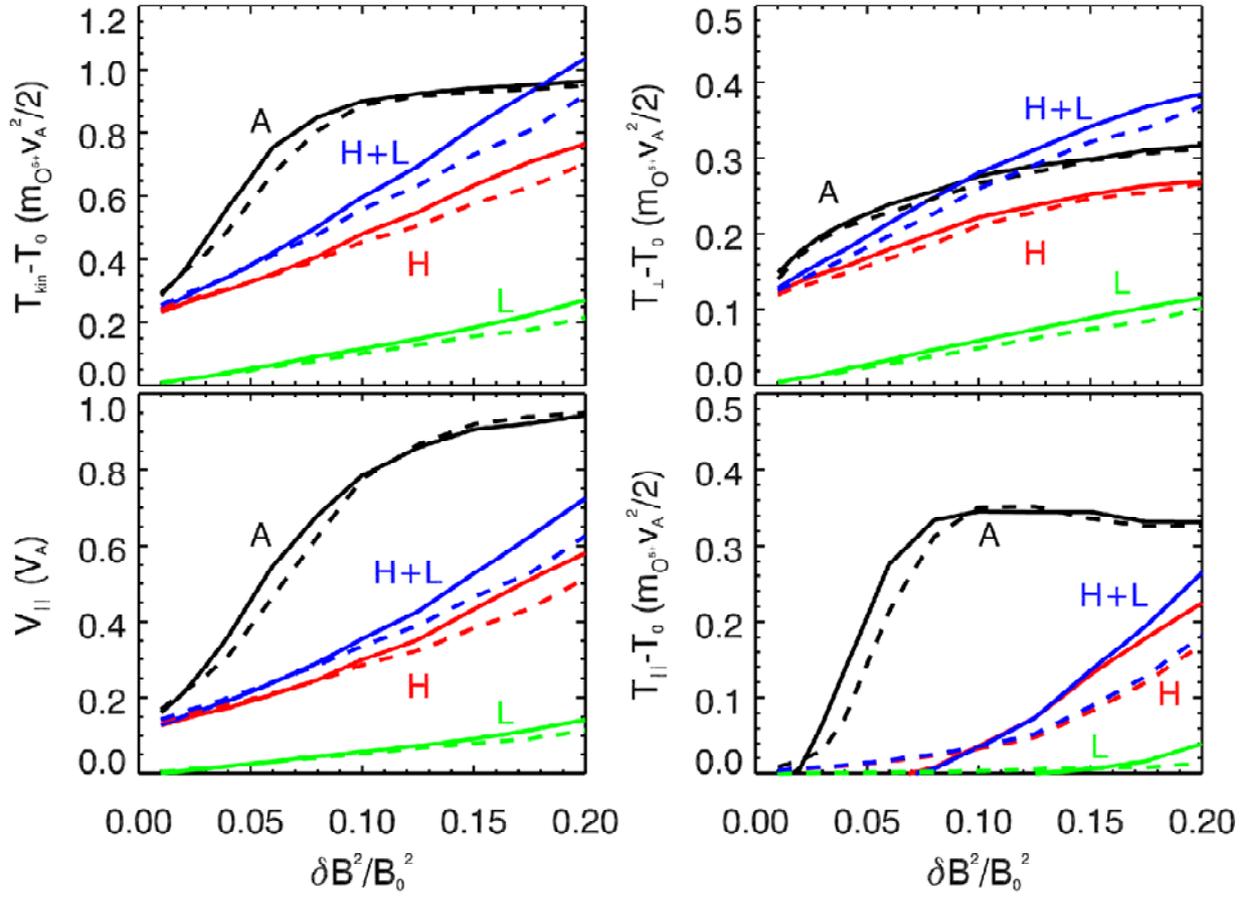

**Figure 12** Variation of the ion kinetic temperature and average parallel speed with the normalized wave energy density $\delta B^2 / B_0^2$ at the end of the simulation time $\Omega_p t = 20000$, where $\delta B^2 / B_0^2$ is the value of the whole wave spectrum in the frequency range $[0.01, 0.4]\Omega_p$. The black lines, red lines and green lines in figure 11 represent the results for Cases A, H and L, respectively. The blue lines are the sum of values obtained in Case H and Case L separately. The results for plasma $\beta = 0.1$ and 0.01 are illustrated by the solid lines and dash lines.